\documentclass[10pt]{article}
\usepackage[english]{babel}
\usepackage{graphics}
\usepackage{graphicx}
\usepackage{amsmath}
\usepackage{amsfonts}
\usepackage{amssymb}%
\begin{document}
\def\beq{\begin{equation}}
\def\eeq{\end{equation}}
\def\bear{\begin{eqnarray}}
\def\ear{\end{eqnarray}}
\newcommand{\Eq}[1]{Eq.\,(\ref{#1})}
\newcommand{\sect}[1]{Sec.\,#1}
\newcommand{\Ref}[1]{Ref.\,\cite{#1}}
\newcommand{\Refs}[1]{Refs.\,\cite{#1}}
\def\s{{\,\rm s}}
\def\g{{\,\rm g}}
\def\eV{\,{\rm eV}}
\def\keV{\,{\rm keV}}
\def\MeV{\,{\rm MeV}}
\def\GeV{\,{\rm GeV}}
\def\TeV{\,{\rm TeV}}
\def\sv{\left<\sigma v\right>}
\def\({\left(}
\def\){\right)}
\def\cm{{\,\rm cm}}
\def\K{{\,\rm K}}
%\begin{verbatim}
\title{Project of Virtual Institute of Astroparticle Physics}
\author{Maxim Yu. Khlopov$^{1,2,3}$\thanks{Maxim.Khlopov@roma1.infn.it}}
\date{}
\maketitle

\begin{center}
{\sl\small $^{1}$ Centre for Cosmoparticle Physics "Cosmion"\\
Moscow, 125047,
Miusskaya pl. 4 \\
$^{2}$ Moscow State Engineering Physics Institute, \\
Kashirskoe Sh., 31, Moscow 115409, Russia, and \\
$^{3}$ APC laboratory 10, rue Alice Domon et L\'eonie Duquet \\75205 Paris
Cedex 13, France}
\end{center}

\begin{abstract}
Studies in astroparticle physics are actively developed all over the world. It is clear that the effectiveness of the work depends strongly on the information exchange rate and on the overall coordination of this activity. An international forum, be it virtual, which can join all the groups and coordinate their efforts would give a boost to this cooperation. Particularly this is important for isolated scientific groups and scientists from small countries which can contribute a lot to this work being a part of the large international collaboration. Objectives, instruments and structure of proposed Virtual Instutute of Astroparticle Physics are discussed.
\end{abstract}

\section{Introduction}
Studies in astroparticle physics link astrophysics, cosmology and particle physics and involve hundreds of scientific groups linked by regional networks (like ASPERA/ApPEC \cite{aspera}) and national centres. The exciting progress in precision cosmology (WMAP and PLANCK missions), in gravitational wave astronomy (VIRGO, GEO, LIGO and  LISA), new cosmic-ray (e.g. PAMELA, GLAST, AMS, HESS, MAGIC, AUGER, ANTARES, NEMO, NESTOR, IceCUBE and KM3) and accelerator experiments (Tevatron and LHC) promise large amount of a new information and discoveries in coming years. The results of these experiments have a direct impact on the development of astroparticle models which also give direct feed-back to the running experiments for search of new effects or particles following from these models. Theoretical analysis of these results will have impact on the fundamental knowledge on the structure of microworld and Universe and on the basic, still unknown, physical laws of Nature (see e.g. \cite{book} for review). 

It is clear that the effectiveness of the work depends strongly on the number of groups involved in this activity, on the information exchange rate and on the overall coordination. 
An international forum, be it virtual, which can join all the groups and coordinate their efforts would give a boost to this cooperation. Particularly this is important for isolated scientific groups and scientists from small countries which can contribute a lot to this work being a part of the large international collaboration. 
A good example of such kind of structure created in 2002 is an International Virtual Observatory \cite{IVOA}. It has demonstrated the work effectiveness and fruitful cooperation of many organisations all over the world. Problems of Virtual Laboratories were recently discussed in \cite{Hut:2007ke}.

\section{Objectives and instruments}
In this proposal it is suggested to organise a Virtual Institute of Astroparticle Physics (VIA), which can play the role of such unifying and coordinating structure for astroparticle physics. 
The activity of the Institute will take place on its website and forum in a form of regular communications between different groups and scientists, working in different scientific fields and parts of the world. This activity would easily allow finding mutual interest and organizing task forces for different scientific topics of astroparticle physics. It can help in the elaboration of strategy of experimental particle, nuclear, astrophysical and cosmological studies as well as in proper analysis of experimental data.
\begin{itemize}
\item[$\bullet$]  Virtual seminars will be experimented, where individual researchers will be able to "attend" from their post by using videoconferencing software (adobe, EVO, marratech etc).  Maximum interactivity will be sought. 
\item[$\bullet$]	The Institute would consider the involvement of young physicists into its activity as one of the most important aspect of the work. It will open the possibility for young people to find scientific direction and even topics of study with possible supervisor. For this purpose the site would provide virtual lectures, and student courses of lectures and training for young scientists from all over the world. 
\item[$\bullet$]	Collaborative spaces (e.g. of the WIKI type) will be created where the exchange of work on specific items will be secured for groups of co-authors. 
\item[$\bullet$]	Public spaces will also be available, where the main results of the domain will be presented, including links involving scientists in studies of particular problems.
\item[$\bullet$]	Databases of .ppt presentations (see the relevant POSbase software \cite{POS}) at topical conferences and schools will make it a source of presentations on different topics.
\end{itemize}
\section{State of the art}
\begin{figure}
    \begin{center}
        \includegraphics[scale=0.7]{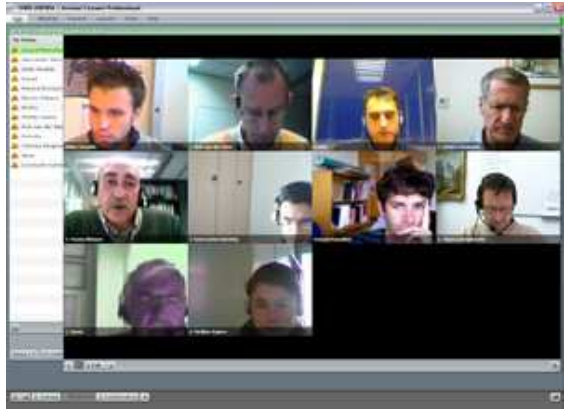}
        \caption{Adobe videoconference Paris-Lyon-Geneve-Hamburg-Moscow-Amsterdam}
        \label{adobe}
    \end{center}
\end{figure}

First tests of VIA system involved:
\begin{itemize}
\item[$\bullet$] Regular Internet EVO course "Introduction to
cosmoparticle physics" by Maxim Khlopov for 5th
year students of Moscow Engineering and Physics
Institute during academic 2006/2007 year.
\item[$\bullet$] On 17 May 2007 this course included Lecture by Prof.
Francesco Fidecaro (see Fig.\ref{Fidecaro}) on gravitational wave detectors.

\begin{figure}
    \begin{center}
        \includegraphics[scale=0.7]{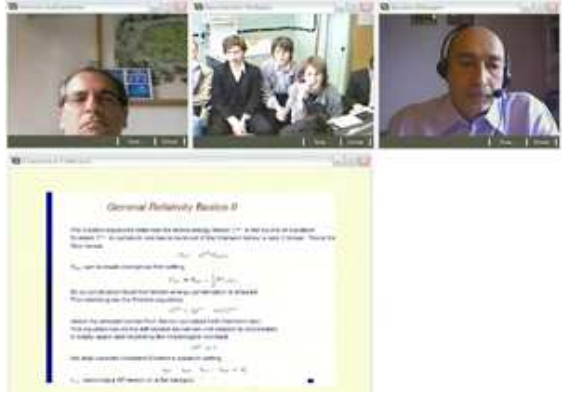}
        \caption{EVO lecture by prof.F.Fidecaro. Slides are demonstrated in Pisa, students attended a room in Moscow, while the two other participants were in Paris}
        \label{Fidecaro}
    \end{center}
\end{figure}
%\begin {figure} \vspace{7cm}\caption{EVO lecture by prof.F.Fidecaro. Slides are demonstrated in Pisa, students attended a room in Moscow, while the two other participants were in Paris} \label{Fidecaro} \end {figure}

\item[$\bullet$] All the students passed online oral Internet exam
on this course with the use of EVO videoconferencing
system.
\item[$\bullet$] On 23 June 2007 lecture by Prof. Hubert Reeves
about problem of light element abundance. 
\item[$\bullet$] On 18 July 2007 lecture by Prof. Etienne Parizot
about latest results of AUGER experiment. 
\item[$\bullet$] Special tests of videoconferencing in case of limited
bandwidth were undertaken in Bled, Slovenia, July 2007 and in
ICTP, Trieste, Italy - august-september 2007
\item[$\bullet$] Studies of different interactive Internet systems
(skype, VRVS, EVO, marratech, adobe, WebEx) in CERN, Switzerland
and ICTP Trieste, Italy - august-december 2007 (see Fig.\ref{adobe})
\end{itemize}
%\begin {figure} \vspace{7cm}\caption{Adobe videoconference Paris-Lyon-Geneve-Hamburg-Moscow-Amsterdam} \label{adobe} \end {figure}
\section{Management}
The structure of the Institute would be minimal to start with. It would include a Scientific Director and a Technical Manager for the web administration. Its scientific content will be supervised by an International Scientific Committee, and its administration by an Administration Council formed by representatives of the purveyors of funds. 
There will be Scientific Sections for each direction of studies with their supervisors and mediators of the corresponding topic of Forum. Each Section should contain a list of lecturers and supervisors for training in the Virtual Institute. 

There are temporarily 4 sections in the Virtual Institute with the following
temporary conveners
\begin{enumerate}
	\item [Section 1.] Cosmology, Physics of Dark Matter and Dark Energy\\ John Ellis, Pierre Binetruy
	\item [Section 2.] High Energy Universe - New Physics in Space\\ Martin Pohl, Angela Olinto
	\item [Section 3.] Neutrino Physics and Astrophysics\\ Stavros Katsanevas, Christian Spieering
	\item [Section 4.] Gravitational experiments and effects\\ Francesco Fidecaro
\end{enumerate}

The communication language will be English but it is assumed that this site would host multi-lingual content beyond that (French, Italian, Spanish, German, Russian…).

\section{\label{Foundation} Foundation}
The efficiency of International activity of Laboratory very strongly depends on the participation of the widely recognized organizations, whose role at the beginning can be limited by some financial support of the WEB designers and scientific and web management. Organization should have an open structure which will evolve in time attracting more and more participants. At present, the first tests of the activity are supported as a pilot project of ApPEC/ASPERA and the information on the program of VIA lectures, which will start in January 2008 can be found in \cite{aspera}.

\section*{Acknowledgements}
 I am grateful to S.Katsanevas for permanent stimulating support, to P.Binetruy, J.Ellis, F.Fidecaro, M.Pohl for help in development of the project, to K.Belotsky and K.Shibaev for assistance in educational tests and to D.Rouable for help in technical realization.

%\end{verbatim}

\end{document}